\documentclass[11pt]{article}
\usepackage{sigsam,amsmath,longtable,tikz}

\issue{}%Vol.xx, No.xx, Issue xxx, month 201x}
\articlehead{}%ISSAC 2025 abstracts}
\titlehead{}
\authorhead{}
\def\<#1>{\langle#1\rangle}

\let\set\mathbb

\begin{document}

\title{Consequences of the Moosbauer-Poole Algorithms}

\author{%
  Manuel Kauers\footnote{Supported by the Austrian FWF grants 10.55776/PAT8258123, 10.55776/I6130, and 10.55776/PAT9952223.}\quad
  and\quad Isaac Wood\footnote{Supported by the Austrian FWF grants 10.55776/PAT8258123}\\
  Institute for Algebra $\cdot$ Johannes Kepler University $\cdot$ Linz, Austria\\
  \url{manuel.kauers@jku.at} $\cdot$ \url{isaac.wood@jku.at}
}

\date{}

\maketitle

\begin{abstract}
  Moosbauer and Poole have recently shown that the multiplication of two $5\times 5$ matrices
  requires no more than 93 multiplications in the (possibly non-commutative) coefficient ring, and
  that the multiplication of two $6\times 6$ matrices requires no more than 153 multiplications.
  Taking these multiplication schemes as starting points, we found improved matrix multiplication
  schemes for various rectangular matrix formats using a flip graph search.
\end{abstract}

More than half a century after the discovery of Strassen's algorithm, we still
do not know how many multiplications in the ground ring are necessary for
multiplying an $n\times m$ matrix with an $m\times p$ matrix. The question is
not only open for asymptotically large matrices but for almost every specific
matrix format. The number of multiplications required by a particular matrix
multiplication scheme is called the \emph{rank} of the scheme. For example,
the rank of the standard algorithm for multiplying $n\times m$ matrices with
$m\times p$ matrices is $nmp$, and the rank of Strassen's algorithm for multiplying
two $2\times 2$ matrices is~$7$, one less than the rank of the standard algorithm. 

One of several known techniques to search for multiplication schemes of low rank
is the flip graph search~\cite{kauers23f,moosbauer23,arai24,kauers24c}, which
takes a known matrix multiplication scheme for a certain format as input and
performs a sequence of operations on it with the goal of obtaining a variant
from which a multiplication can be eliminated. For several small matrix formats,
the multiplication schemes with the smallest rank known today were found with
this method. Most recently, Moosbauer and Poole~\cite{moosbauer25} used a
variant of the flip graph method that takes symmetries into account in order to
obtain improvements for the formats $(n,m,p)=(5,5,5)$ and
$(n,m,p)=(6,6,6)$. Their schemes have rank 93 and 153, respectively.

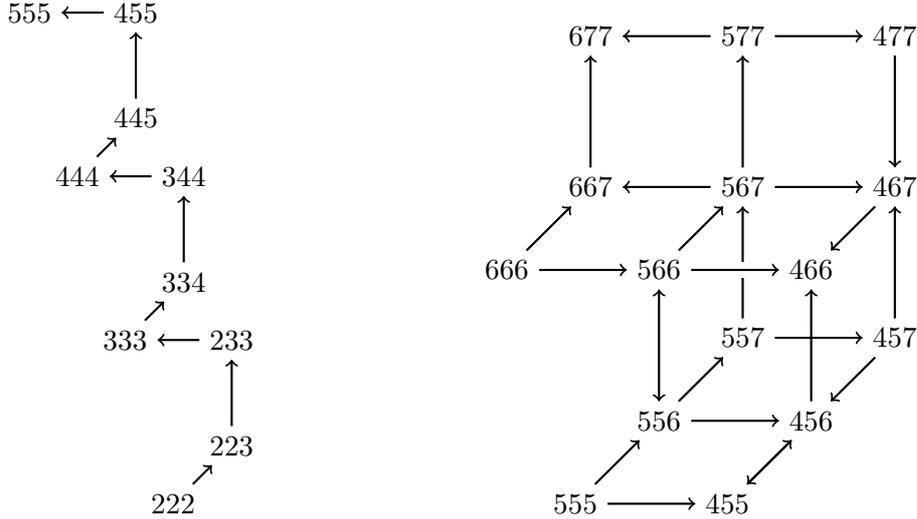
\begin{figure}
\begin{center}
  \begin{tikzpicture}[x={(-1cm,0cm)},z={(.55cm,.55cm)},y={(0cm,1cm)},scale=1.4]
    \node (222) at (2,2,2) {222};
    \node (223) at (2,2,3) {223};
    \node (233) at (2,3,3) {233};
    \node (333) at (3,3,3) {333};
    \node (334) at (3,3,4) {334};
    \node (344) at (3,4,4) {344};
    \node (444) at (4,4,4) {444};
    \node (445) at (4,4,5) {445};
    \node (455) at (4,5,5) {455};
    \node (555) at (5,5,5) {555};
    \draw[->,thick] (222)--(223);
    \draw[->,thick] (223)--(233);
    \draw[->,thick] (233)--(333);
    \draw[->,thick] (333)--(334);
    \draw[->,thick] (334)--(344);
    \draw[->,thick] (344)--(444);
    \draw[->,thick] (444)--(445);
    \draw[->,thick] (445)--(455);
    \draw[->,thick] (455)--(555);
  \end{tikzpicture}\hfil
  \begin{tikzpicture}[x={(-1cm,0cm)},z={(.55cm,.55cm)},y={(0cm,1cm)},scale=2]    
    \node (455) at (4,5,5) {455};
    \node (456) at (4,5,6) {456};
    \node (457) at (4,5,7) {457};
    \node (466) at (4,6,6) {466};
    \node (467) at (4,6,7) {467};
    \node (477) at (4,7,7) {477};
    \node (555) at (5,5,5) {555};
    \node (556) at (5,5,6) {556};
    \node (557) at (5,5,7) {557};
    \node (566) at (5,6,6) {566};
    \node (567) at (5,6,7) {567};
    \node (577) at (5,7,7) {577};
    \node (666) at (6,6,6) {666};
    \node (667) at (6,6,7) {667};
    \node (677) at (6,7,7) {677};
    \draw (455)--(456)--(457) (466)--(467)--(477) (456)--(466) (457)--(467);
    \draw (555)--(556)--(557) (556)--(557) (556)--(566)--(567)--(577) (557)--(567);
    \draw (455)--(555) (456)--(556) (466)--(566) (457)--(557) (467)--(567) (477)--(577);
    \draw (666)--(667)--(677) (566)--(666) (567)--(667) (577)--(677);
    \begin{scope}[thick,->]
      \draw (666)--(566);
      \draw[<->] (566)--(556);
      \draw (666)--(667);
      \draw (667)--(677);
      \draw (555)--(556);
      \draw (566)--(567);
      \draw (567)--(577);
      \draw (567)--(667);
      \draw (556)--(557);
      \draw (557)--(567);
      \draw (555)--(455);
      \draw (556)--(456);
      \draw (557)--(457);
      \fill[white](566)++(3mm,-.5mm) rectangle ++(4mm,1mm);
      \draw (566)--(466);
      \draw (567)--(467);
      \draw (577)--(477);
      \draw (577)--(677);
      \draw (456)--(466);
      \draw (477)--(467);
      \draw (457)--(467);
      \draw (457)--(456);
      \draw (455)--(456);
      \draw (456)--(455);
      \draw (467)--(466);
    \end{scope}
  \end{tikzpicture}
\end{center}
\caption{Left: Arai et al. obtained a good scheme for $(5,5,5)$ starting from a scheme of size $(2,2,2)$ by following the
  depicted path. 
  Right: We performed flip graph searches for various formats, using starting points obtained along the indicated
  arrows from the schemes found by Moosbauer and Poole for $(5,5,5)$ and $(6,6,6)$. For some formats, we tried
  several paths. 
}\label{fig:1}
\end{figure}

For the present work, it is not necessary to know in detail how the flip graph
search works. It suffices to know that the search can start from an arbitrary
correct multiplication scheme for a certain format and then tries to eliminate
multiplications from it. A natural start point for the search is the standard
algorithm. While this choice works well for small matrix formats, it was already
observed by Kauers and Moosbauer~\cite{kauers23f} that it is not always the best choice. For
$(5,5,5)$, they obtained a better result using a scheme found by
AlphaTensor~\cite{FBH+:Dfmm} rather than the standard algorithm. Arai et al.~\cite{arai24} obtained
an even better result for $(5,5,5)$ using an incremental approach. They
construct their starting points for a given format $(n,m,k)$ from good schemes
they found for smaller formats. This works because matrices can be multiplied
blockwise. For example, a multiplication scheme of format $(3,3,4)$ can be
obtained by patching together a scheme of format $(3,3,3)$ and a scheme of
format $(3,3,1)$:
\[
\begin{pmatrix}
  a_{1,1} & a_{1,2} & a_{1,3} \\
  a_{2,1} & a_{2,2} & a_{2,3} \\
  a_{3,1} & a_{3,2} & a_{3,3}
\end{pmatrix}\cdot
\begin{pmatrix}
  b_{1,1} & b_{1,2} & b_{1,3} & b_{1,4}\\
  b_{2,1} & b_{2,2} & b_{2,3} & b_{2,4}\\
  b_{3,1} & b_{3,2} & b_{3,3}\smash{\rlap{\kern.5em\rule[-.5ex]{.7pt}{3.5em}}} & b_{3,4}
\end{pmatrix}=
\begin{pmatrix}
  c_{1,1} & c_{1,2} & c_{1,3} & c_{1,4}\\
  c_{2,1} & c_{2,2} & c_{2,3} & c_{2,4}\\
  c_{3,1} & c_{3,2} & c_{3,3}\smash{\rlap{\kern.5em\rule[-.5ex]{.7pt}{3.5em}}} & c_{3,4}
\end{pmatrix}.
\]
Arai et al. obtained their improvement for $(5,5,5)$ by a sequence of flip graph searches for the formats
$(2,2,2)$, $(2,2,3)$, $(2,3,3)$, $(3,3,3)$, $(3,3,4)$, $(3,4,4)$, $(4,4,4)$, $(4,4,5)$, $(4,5,5)$,
$(5,5,5)$, always taking starting points obtained by extending the results of the search of the
previous format (cf. Fig.~\ref{fig:1}, left).

In the same spirit, we explore formats in the vicinity of $(5,5,5)$, taking the Moosbauer-Poole algorithms as
basis. In addition to extending known schemes to schemes of larger format, we can also use a known
scheme to construct promising starting points for smaller formats, by simply setting some variables
to zero. For example, every scheme of format $(3,3,3)$ contains a scheme of format $(2,3,3)$ via
\[
\begin{pmatrix}
  a_{1,1} & a_{1,2} & a_{1,3} \\
  a_{2,1} & a_{2,2} & a_{2,3} \\
  0 & 0 & 0 
\end{pmatrix}\cdot
\begin{pmatrix}
  b_{1,1} & b_{1,2} & b_{1,3} \\
  b_{2,1} & b_{2,2} & b_{2,3} \\
  b_{3,1} & b_{3,2} & b_{3,3} 
\end{pmatrix}=
\begin{pmatrix}
  c_{1,1} & c_{1,2} & c_{1,3} \\
  c_{2,1} & c_{2,2} & c_{2,3} \\
  0 & 0 & 0 
\end{pmatrix}.
\]
Altogether, we considered the formats shown in Fig.~\ref{fig:1} on the right. The arrows indicate from
where the starting points of the flip graph search were obtained.
Because of the symmetry of the matrix multiplication tensor, it suffices to consider
formats $(n,m,k)$ with $n\leq m\leq k$.

As an example, consider the format $(5,6,7)$. As illustrated in Fig.~\ref{fig:1}, we
approached this format from $(5,6,6)$ and from $(5,5,7)$. For the format
$(5,6,6)$ we have a multiplication scheme of rank~130. Extending this scheme to
a scheme for $(5,6,7)$ costs $5\times6=30$ additional multiplications. So the
starting point in this case is a scheme of rank~160. A flip graph search brings
this down to~150.  For $(5,5,7)$ we have a scheme of rank 127, which extends to
a scheme for $(5,6,7)$ of rank~162.  In this case, the flip graph search could
bring only down to~152, so the first attack yielded the better result.

Some arrows are computationally more expensive than others, for reasons that are
unclear. On the average, we have spent roughly one day of computation time on a
machine with 100 cores for each arrow. 

The results of all our computations are summarized in the table below. It turns
out that there are improvements for all formats except for $(4,6,6)$ and
$(6,7,7)$, and in some cases the improvement is considerable. The column
``previous record'' refers to the best known rank according to Sedoglavic's
table~\cite{fastmm} as of April 2025. These ranks refer to ground fields of
characteristic zero.

Our flip graph search first produces schemes for the ground field~$\set Z_2$.
In order to translate them to schemes with integer coefficients, we attempted to
apply Hensel lifting, and most cases, at least one of the schemes with the
smallest rank over $\set Z_2$ could be lifted to a scheme with integer
coefficients. For $(4,5,5)$, we only found schemes of rank~74 over $\set Z_2$
which cannot be lifted, like Arai et al.~\cite{arai24}. Also none of the schemes
of rank~75 we found for this format can be lifted, but some of rank~76 can. For
$(6,7,7)$, the schemes we found of rank 221 can also not be lifted, but in this
case we did not spend a lot of effort finding schemes with integer coefficients
because 221 is larger than the best known rank.

\begin{center}
  \begin{tabular}{c|ccc}
    format & naive & previous & our \\
           & rank & record & rank  \\\hline
    $(4,5,5)$ & 100 & 76 & 76  \\
    $(4,5,6)$ & 120 & 93 & \textbf{90} \\
    $(4,5,7)$ & 140 & 109 & \textbf{104} \\
    $(4,6,6)$ & 144 & 105 & 106 \\
    $(5,5,6)$ & 150 & 116 & \textbf{110} \\
    $(4,6,7)$ & 168 & 125 & \textbf{123} \\
    $(5,5,7)$ & 175 & 133 & \textbf{127} \\
    $(5,6,6)$ & 180 & 137 & \textbf{130} \\
    $(4,7,7)$ & 196 & 147 & \textbf{144} \\
    $(5,6,7)$ & 210 & 159 & \textbf{150} \\
    $(5,7,7)$ & 245 & 185 & \textbf{176} \\
    $(6,6,7)$ & 252 & 185 & \textbf{183} \\
    $(6,7,7)$ & 294 & 215 & 221\rlap{ (over~$\set Z_2$)}
  \end{tabular}
\end{center}

The multiplication schemes announced here are publicly available at \url{https://github.com/mkauers/matrix-multiplication}.

\bibliographystyle{plain}
\bibliography{bib}

\end{document}